\documentclass[12pt,osajnl2,preprint,showpacs,superscriptaddress]{revtex4}  

\begin{document}

\title{Impact of Decoherence on Internal State Cooling using Optical Frequency Combs }

\author{ S.A. Malinovskaya}
\affiliation{Department of Physics and Engineering Physics, Stevens Institute of
Technology, Hoboken, NJ 07030}
\email{smalinov@stevens.edu}
\author{S.L. Horton}
\affiliation{Department of Physics and Engineering Physics, Stevens Institute of
Technology, Hoboken, NJ 07030}

\begin{abstract}
We discuss femtosecond Raman type techniques to control molecular vibrations, which can be implemented for internal state cooling from Feshbach states with the use of optical frequency combs with and without modulation. The technique makes use of multiple two-photon
resonances induced by optical frequencies present in the comb. It
provides us with a useful tool to study the details of molecular
dynamics at ultracold temperatures. In our theoretical model we take into account decoherence in the form of spontaneous emission and collisional dephasing in order to ascertain an accurate model of the population transfer in the three-level system. We analyze the effects of odd and even chirps of the optical frequency comb in the form of sine and cosine functions on the population transfer. We compare the effects of these chirps to the results attained with the standard optical frequency comb to see if they increase the population transfer to the final deeply bound state in the presence of decoherence. We also analyze the inherent phase relation that takes place owing to collisional dephasing between molecules in each of the states. This ability to control the rovibrational states of a molecule with an optical frequency comb enables us to create a deeply bound ultracold polar molecules from the Feshbach state.
\end{abstract}


\maketitle

Decoherence at ultracold temperatures is the subject of particular
interest and importance in light of the development of the methods
to manipulate ultracold gases, create and control ultracold
molecules \cite{De11,Qi10}, study ultracold collisions and chemical reactions \cite{Ni10}.
Decoherence is inherently present in ultracold dynamics and we
 study it semiclassically within the process of creation of diatomic KRb
molecules from Feshbach, weakly bound states. Experimentally ultracold polar KRb molecules were obtained using
stimulated Raman adiabatic passage (STIRAP), \cite{Ni08}. As a viable substitute to the STIRAP process, we created a method which makes use of  optical
frequency combs to perform two-photon resonances and transfer population
in a stepwise manner from the Feshbach state to the ground,
ultracold state.
The implementation of an optical frequency comb (OFC) may be beneficial
owing to its intrinsic ability to address the manifold of excited states simultaneously. We
make use of a standard OFC, and the one with the sinusoidal phase
modulation to induce two-photon Raman transitions.  Our theory demonstrates that the impact of
decoherence may be minimized by implementing the sinusoidally
modulated OFC.

\begin{figure}
\centerline{\includegraphics[width=5cm]{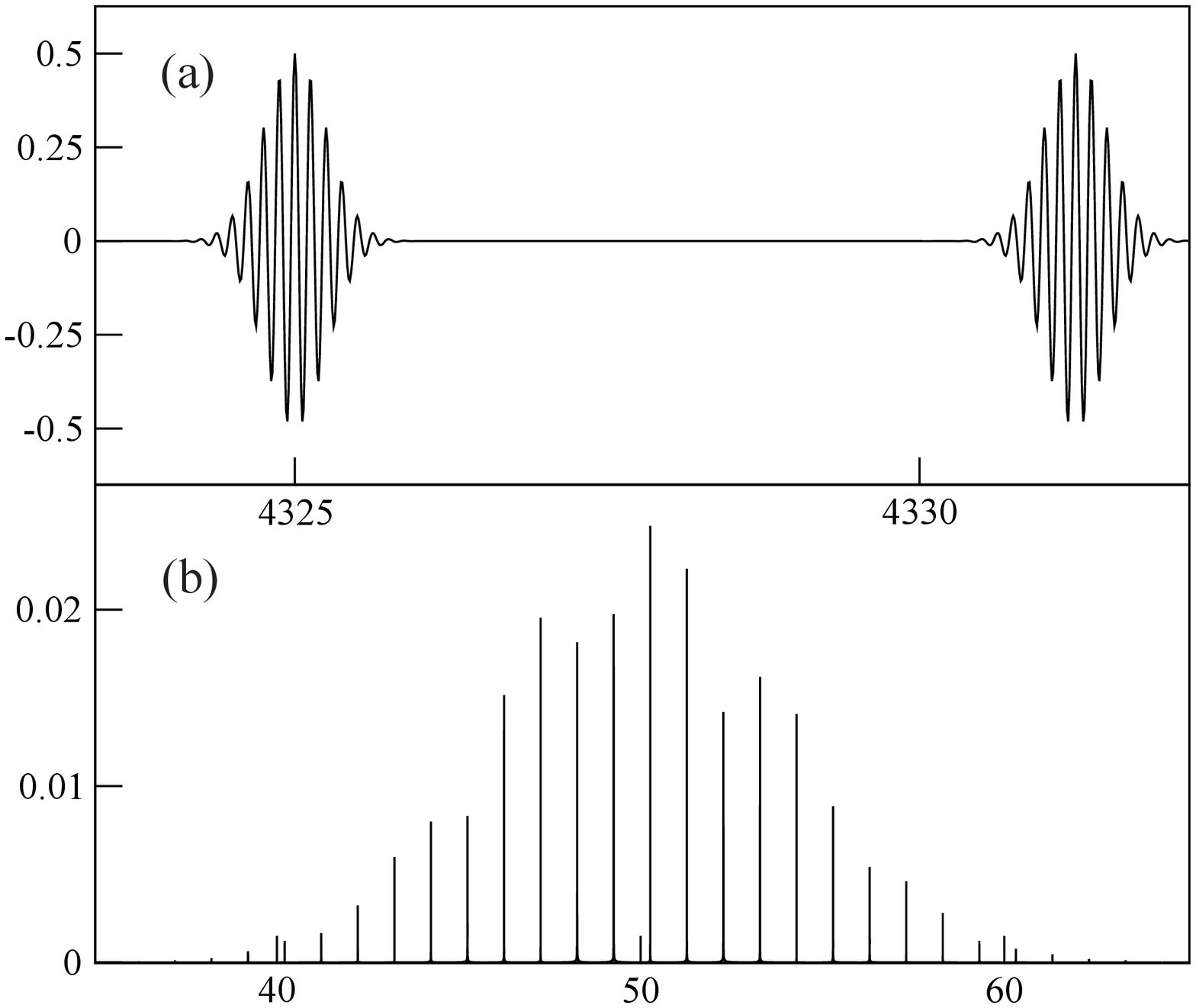}
\hspace{1.cm}\includegraphics[width=5cm]{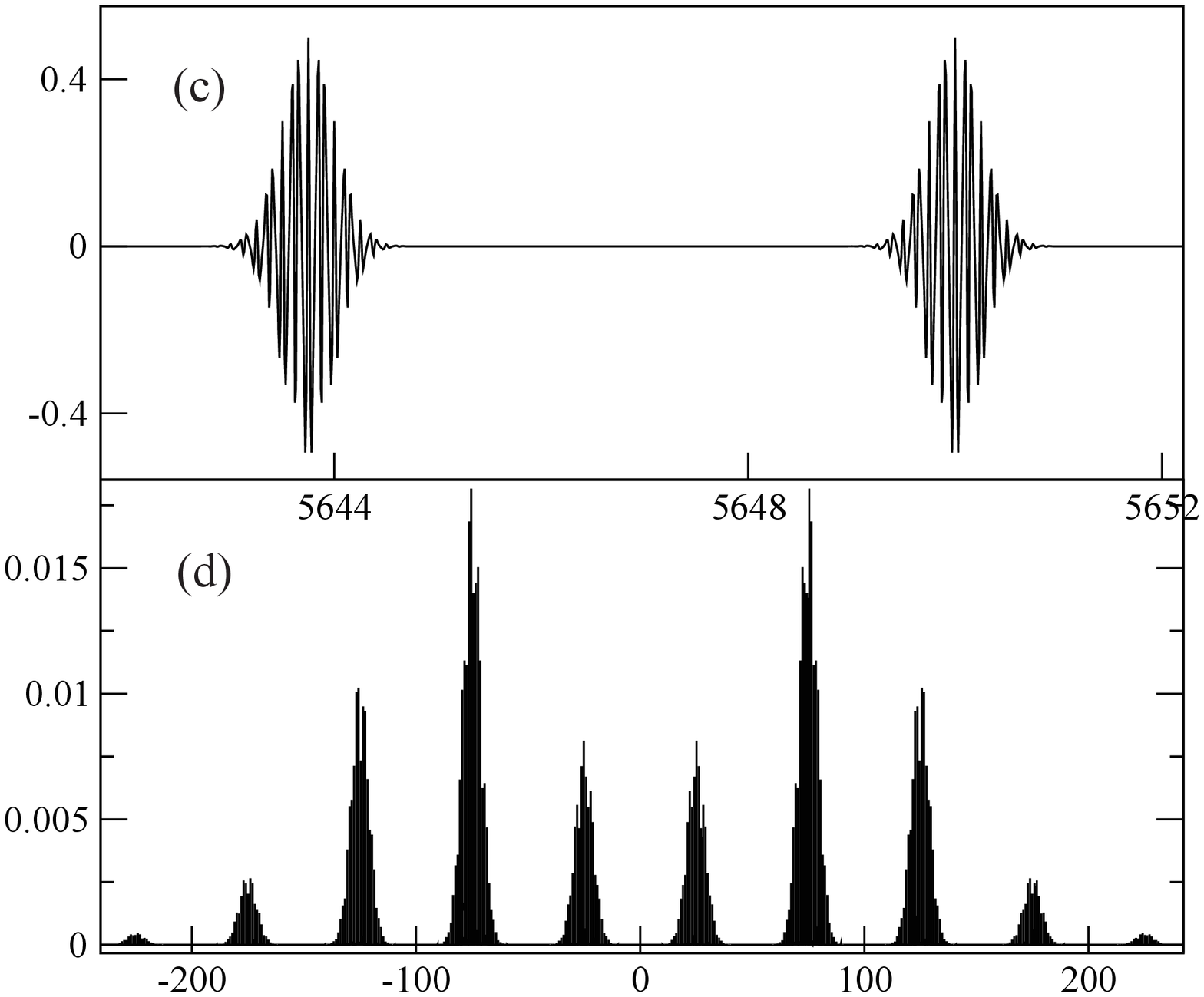}}
\caption{(a) A qualitative figure of the pulse train in Eq.(\ref{FC0} and (b) the respective frequency comb.
(c) The sinusoidally modulated pulse train as in Eq.(\ref{FC})
and (d) the frequency comb with sets' uniform spacing determined by $\Omega$ and
spacing between the modes precisely equal to $2\pi
T^{-1}.$
}\label{comb2}
\end{figure}

 The standard OFC is generated by the phase locked
pulse train of the form
\begin{eqnarray}
\label{FC0}&E(t)=\Sigma_{k=0}^{N-1}
E_0\exp(-(t-kT)^2/(2\tau^2))\cos{(\omega_L t - k \omega_L T+\phi)}.
\end{eqnarray}
Eq.(\ref{FC0}) gives a periodic envelope of the field which
oscillates with an optical carrier frequency $\omega_L$, and $k$ is an integer number. The pulse train period $T$ is much greater than $\tau$, where $\tau$ is
an individual pulse duration. A strictly periodic envelope function
can be expressed as a spectrum by the Fourier series
$\Sigma_{q=-\infty}^{+\infty} A_q exp{(-i(\omega_L + q \omega_r) t)}
+c.c.$, where $\omega_r=2\pi T^{-1}$ is the pulse repetition rate
 and $A_q$ are spectral intensities that do not depend on
time. The spectrum is a comb of laser frequencies precisely spaced
by the pulse repetition rate $\omega_r$. A qualitative picture of
the temporal field and the frequency comb are shown on
Fig.(\ref{comb2} a,b).
The period of the pulse train may be within a 10 ns time
scale, e.g. \cite{St08}, it determines the spacing between modes in the frequency
comb to be 100 MHz. The
Ti:Saphfire laser gives rise to about $10^6$ modes in the comb.

We study the phase variation across an individual pulse in the form
of the {\em sin} function
\begin{equation}
\label{FC}E(t,z)=1/2 \Sigma_{k=0}^{N-1} E_0
\exp(-(t-kT)^2/(2\tau^2)) \cos {(\omega_L (t-kT) + \Phi_0 \sin
(\Omega (t-kT)) +\phi)}.
\end{equation}
Here the $\Phi$ is the modulation amplitude and the $\Omega$ is the modulation frequency.
The time-dependent phase of the field in the form of a {\em sin}
function brings additional peaks to the frequency comb. Laser
frequency $\omega_L$ determines the center of the frequency comb,
while $\Omega$ determines the uniform spacing of the individual sets
of modes. Within each set of modes, the uniform spacing between the
modes is determined by $2\pi T^{-1}$. Eq.(\ref{FC}) gives the pulse
train and the intensity spectrum qualitatively shown in Fig.(\ref{comb2} c,d).
Increasing $\Omega$
broadens the frequency comb region, while $\Phi_0$ determines the
amplitude of the comb's peaks.
Frequency comb peaks are narrower for longer propagation time of the
pulse train, that is $1/t_{total}$ determines the bandwidth of an
individual mode.

A sinusoidal modulation at the MHz frequency of the beam of a cw ring dye laser was implemented in \cite{Ha81} to study absorption resonances in $I_2$ with high precision. Here we assume modulation in THz region for which the approach which makes use of an enhancement cavity may be applied \cite{Su08}.

\begin{figure}
\centerline{\includegraphics[height=7cm]{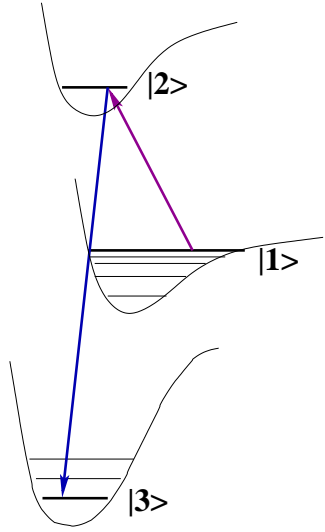}}
\caption{A three-level $\Lambda$ system modeling the Feshbach state, the excited state manifold and the final state of molecules interacting with a phase locked pulse 
}\label{scheme_FC_decoh}
\end{figure}

We consider a three-level $\Lambda$ system interacting with a frequency comb do describe semiclassically the dynamics of the Feshbach molecules interacting with radiation resulting in their stepwise transformation into the ultracold molecules. The $\Lambda$ system is formed by the energy level that include the Feshbach state, the excited, transitional state and the final, ultracold state, Fig(\ref{scheme_FC_decoh}).

For the description of the system dynamics we refer to the Leouville-von
Neumann equation which provides a
set of coupled differential equations for density matrix elements

\begin{eqnarray} \label{LvN}
&\dot{\rho}_{11}=2Im[H_{12}\rho_{21}+H_{13}\rho_{31}] \nonumber \\
&\dot{\rho}_{22}=2Im[H_{21}\rho_{12}+H_{23}\rho_{32}] \nonumber \\
&\dot{\rho}_{33}=2Im[H_{31}\rho_{13}+H_{32}\rho_{23}] \nonumber \\
&\dot{\rho}_{12}=-i H_{12}(\rho_{22}-\rho_{11})-iH_{13}\rho_{32}+iH_{32}\rho_{13}  \\
&\dot{\rho}_{13}=-i H_{13}(\rho_{33}-\rho_{11})-iH_{12}\rho_{23}+iH_{23}\rho_{12} \nonumber \\
&\dot{\rho}_{23}=-i H_{23}(\rho_{33}-\rho_{22})-iH_{21}\rho_{13}+iH_{13}\rho_{21}
 \nonumber
\end{eqnarray}

The interaction Hamiltonian used in these equations is written
beyond the rotating wave approximation. Nonzero
Hamiltonian matrix elements in the interaction representation read $H_{ji}=\Omega_R(t-T)
[\exp\{- i ((\omega_L + \omega_{ji})(t-T) + M(t-T))\} + \exp\{ i
((\omega_L - \omega_{ji})(t-T) + M(t-T))\}]$, here, i,j are the indexes of the basis set, $i=1,2$ and $j=i+1$, $M(t-T)= \Phi_0 \sin \Omega (t-T)$ is the phase modulation in a
 single
 pulse which is zero for the standard OFC, and $\Omega_R(t-T)= \Omega_R \exp{(-(t-T)^2/(2\tau^2))}$ is the Rabi frequency, with the peak value $\Omega_R$.
The decoherence effects caused by spontaneous emission and collisions are taken into account through the reduced density matrix elements
\begin{equation}
\begin{array}{ccc}
\dot{\rho}_{11})_{sp}=\gamma_{21}\rho_{22}  &
 \dot{\rho}_{12})_{sp,col}=-(
 \frac{\gamma_{21}}{2}+\frac{\gamma_{23}}{2}+\Gamma_{21} )
\rho_{12} & \\
\dot{\rho}_{22})_{sp}=-\gamma_{21}\rho_{22}-\gamma_{23}\rho_{22}&
\dot{\rho}_{13})_{sp,col}=-\Gamma_{31}
\rho_{13} & \\
 \dot{\rho}_{33})_{sp}=\gamma_{23}\rho_{22}  &
\dot{\rho}_{23})_{sp,col}=-(
  \frac{\gamma_{21}}{2}+\frac{\gamma_{23}}{2}+\Gamma_{23} )
\rho_{23}, & \\
\end{array} \label{REDST_DENSITY}
\end{equation}
here $\gamma_{21}, \gamma_{32}$ are spontaneous emission rates from the excited state $|2>$ to the initial state $|1>$ and the final state $|3>$ respectively.
The $\gamma$s may also be attributed to inelastic collision rates.
The collisional dephasing is taken into account by the rates $\Gamma_{21}, \Gamma_{31}, \Gamma_{23}$. The reduced density matrix elements were added to Eqs.(\ref{LvN}) which were solved numerically.

To demonstrate the powerful tool of a modulated OFC to control rovibrational degrees of freedom at ultracold temperatures in the presence of decoherence, we analyze the  impact of different channels individually within our model.
At first, we discuss the case when decoherence is not taken into account and emphasize
similarities and differences in the dynamics induced by the standard and the modulated OFCs. Then,
we turn on the decoherence in our model experiment, and illuminate
the advantages the sinusoidally modulated OFC brings to ultrafast
dynamics by minimizing the effects caused by fast spontaneous
emission and collisions.

\begin{figure}
\vspace{20pt} \centerline{
\includegraphics[width=3in]{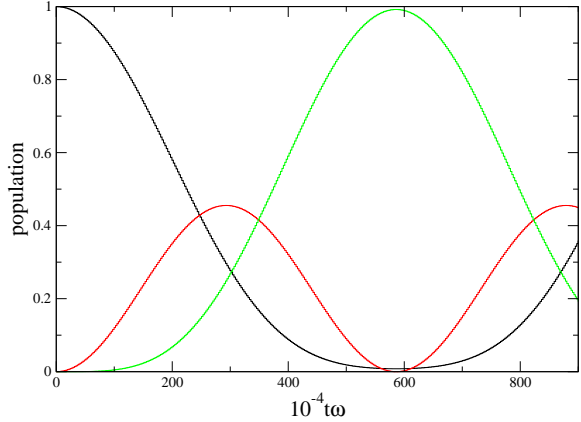}}
\caption{ Population dynamics in the three-level $\Lambda$ system
using an optical frequency comb having
 $f_r$=5 GHz, and zero offset
frequency. Parameters of the pulse train are the carrier frequency
$\omega_L$=434.8 THz, the pulse duration $\tau_0$=3 fs, the peak
Rabi frequency $\Omega_R$=1.26 THz; the system parameters are
$\omega_{21}$=309.3 THz, $\omega_{32}$=434.8 THz, and
$\omega_{31}$=125.5 THz \cite{Ni08}. Black line shows population of the ground
state, red line - population of the transitional state, and green
line - that of the final state. Time is given in the units of
$[\omega^{-1}]$, where $\omega=\omega_{31}$=125.5 THz.
}\label{comb_cooling_Stndrd_AIN0.01_timestp_0.05_Science_res_T_640000}
\end{figure}

The standard OFC and the sinusoidally modulated OFC application to
 a three-level $\Lambda$ system leads to a full
population transfer to the ultracold state from the Feshbach state \cite{Sh10}.
The difference is in the degree of involvement of the intermediate
state into dynamics. The standard OFC
populates substantially the excited, transitional state; its
population may rise up to $50\%$ depending on whether the carrier
frequency of the pulse train is in resonance with the
transition from the excited to ultracold state or is detuned. The
results for the case of the one-photon resonance condition are presented on Fig.(\ref{comb_cooling_Stndrd_AIN0.01_timestp_0.05_Science_res_T_640000}).

\begin{figure}
\vspace{20pt} \centerline{
\includegraphics[width=3in]{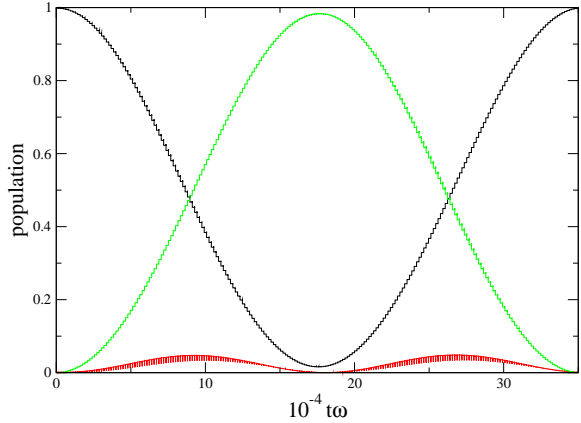}}
\caption{Population transfer in the three-level $\Lambda$ system,
achieved via the resonant Raman transitions using a phase-modulated
optical frequency comb described by Eq.(\ref{FC}).
 The values of the system parameters
are  $\omega_{32}$ = 410.7 THz, $\omega_{21}$ = 340.7 THz, \cite{Sh08}. The carrier frequency $\omega_L$ = $\omega_{32}$, the modulation
frequency $\Omega$ = $\omega_{21}$, the modulation amplitude $\Phi_0=4$
and the peak Rabi frequency $\Omega_R$=70 THz.
  Stepwise, adiabatic accumulation of the
population is observed in state $|3>$, (green), which is the
ultracold KRb state. The population of the Feshbach state $|1>$,
(black), comes gradually to zero, while the excited state manifold
$|2>$, (red), is slightly populated during the transitional time.
Full population transfer is accomplished in 109 pulses. Time is
given in the units of $[\omega^{-1}]$, where $\omega=\omega_{31}$=70
THz. }\label{comb_cooling_AMP4_OM4.9_w_L5.9_frag}
\end{figure}

In contrast to the effect induced by the standard OFC, the sinusoidally modulated
OFC, while performing a stepwise population transfer to the final,
ultracold state, only negligibly populates the excited state. This
dynamics is observed under the condition of the one-photon resonances of the
carrier and the modulation frequencies with the Feshbach-to-excited
state and the excited-to-ultracold state transitional frequencies respectively. The
dynamics is presented in Fig.(\ref{comb_cooling_AMP4_OM4.9_w_L5.9_frag}).

Before discussing the impact of decoherence, let us point our attention to the dependence of the dynamics on the parity of the applied chirp of the field. The implementation of the odd and even chirps may result in a substantially different response of the system and the quantum yield. We investigated the effect of the sine and cosine modulation within the three-level model and revealed an significant dependence of the population dynamics on the parity of the phase modulation. A sinusoidal modulation across an individual pulse in a phase locked pulse train leads to a stepwise adiabatic population transfer as described above, while a cosine modulation induces population jumps from the initial to the final state within a much shorter timescale and is accompanied by a significant population of the excited state during the transitional time. The phenomenon strongly depends on the magnitude of the Rabi frequency. According to \cite{Go02} the different results of the sine and cosine chirp are expected. The authors analyzed the Taylor series expansion of the field phase having an arbitrary time dependence and showed that even chirps lead to self-induced transparency, whereas odd chirps to population inversion.

\begin{figure}
\vspace{20pt} \centerline{
\includegraphics[width=3in]{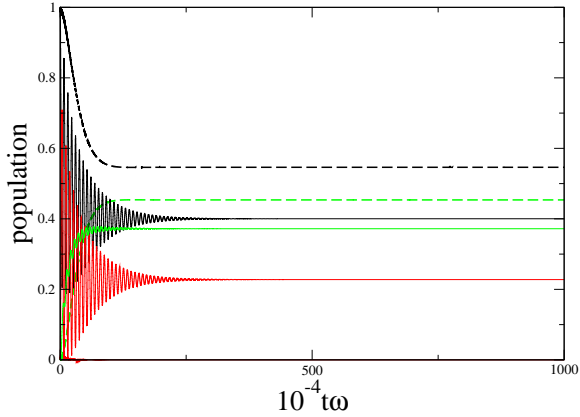}}
\caption{Population dynamics induced by standard optical frequency comb in the presence of spontaneous decay and collisions.
Parameters of the pulse train are the carrier frequency
$\omega_L$=434.8 THz, the pulse duration $\tau_0$=3 fs, the peak
Rabi frequency $\Omega_R$=12.6 THz; the system parameters are
$\omega_{21}$=309.3 THz, $\omega_{32}$=434.8 THz, and
$\omega_{31}$=125.5 THz. Solid lines represent the case of pure collisional dephasing, $\Gamma_{31}=0$, $\Gamma_{21}=\Gamma_{32}=0.001$, dashed lines correspond to the same values of $\Gamma$'s and $\gamma_{21}=\gamma_{32}=0.001$.  }\label{comb_cooling_g210.001_gg320.001_vs_g21zero_gg320.001_3200_scie_stdrt}
\end{figure}

In the presence of fast decoherence, different degree of the involvement of the excited state into population dynamics induced by
 the standard and the sinusoidally modulated OFCs leads to a remarkably different response of the three-level system to the applied field. We decompose the contributions from spontaneous decay and collisional dephasing in our analysis. The time-dependent dynamics induced by the standard OFC in the presence of only collisional dephasing is
presented by solid lines in Fig.(\ref{comb_cooling_g210.001_gg320.001_vs_g21zero_gg320.001_3200_scie_stdrt}).
Here, the quantum yield to the ultracold state is only $38\%$ in the presence of fast collisions. When we add  the spontaneous decay from the excited state, we observe that it reduces the collisional impact and yields to a higher steady state population of the final, ultracold state equal to $45\%$, see Fig.(\ref{comb_cooling_g210.001_gg320.001_vs_g21zero_gg320.001_3200_scie_stdrt}), (dashed lines). The data are presented for the following decoherence parameters: solid lines for $\gamma_{21}=0.$, $\Gamma_{12}=\Gamma_{23}=0.001$; and dashed lines for  $\gamma_{21}=\gamma_{32}=0.001$, and $\Gamma_{12}=\Gamma_{23}=0.001$, in the units of $\omega_{31}$ = 125THz, \cite{Ni08}. The Rabi frequency $\Omega_R$ is equal to 0.1. The interplay of the spontaneous decay rate and the collision dephasing rate may
result in an increase of the quantum yield of ultracold molecules.

\begin{figure}[hbt!]\vspace{0.5cm}
\centerline{
\includegraphics[width=3in]{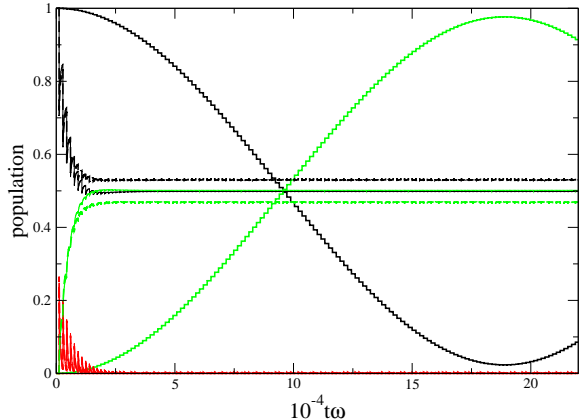}}
\caption{ \small  Population dynamics induced by the modulated optical frequency comb (solid lines) vs the standard one (dashed lines), $\gamma_{21}=\gamma_{32}=0.001$, and $\Gamma_{21}=\Gamma_{32}=0.001$ in $\omega_{31}$ units ($\omega_{31}$ = 125THz). Sinusoidal modulation provides almost full population transfer to the ultracold state, while cosine-modulation leads to a stationary solution with equal population distribution between the Feshbach and ultracold state, thus creating the maximum coherence.}
\label{comb_cooling_Scie_g210.001_gg320.001_3200}
\end{figure}

Next, we demonstrate the effect caused by the sinusoidal modulation across an individual pulse in the phase-locked pulse train. It results in a remarkably high quantum yield to the final, ultracold state under the same decoherence conditions. A comparison is shown on Fig. (\ref{comb_cooling_Scie_g210.001_gg320.001_3200}), where the dashed lines correspond to the standard OFC implementation same as in the previous figure,  solid lines show the system response to the sine modulated OFC. The sinusoidally modulated comb negligibly involves the excited state into transitional dynamics (Fig.(\ref{comb_cooling_AMP4_OM4.9_w_L5.9_frag})), minimizing the effects caused by decoherence. Two horizontal solid lines coinciding at the population equal to 0.5 show the steady state solution when the cosine modulation across an individual pulse in the pulse train was implemented justifying the importance of the parity of the chirp in achieving the desired quantum yield.
Notably, we have uncovered a rigor expression for the relation between collisional dephasing rates in the three-level $\Lambda$ system. Previously it was observed that for certain values of collisional dephasing rates
the received populations at times incorrectly exceeded the initial total population. However, if to look from the physical point of view at several special cases, e.g., when molecules in ultracold configuration collide elastically with the atomic Rb and K at the same rate as the Feshbach molecules, then one would see that the state $|3>$ and $|1>$ amplitudes loose their phase synchronously  and thus the $\Gamma_{31}$ must be zero, while $\Gamma_{21}$ and $\Gamma_{32}$ must be not and be equal to each other. This example demonstrates that the values of collisional dephasing rates are mutually dependent. With the help of the phase diagram, we obtained a general relation between dephasing rates in the three-level $\Lambda$ system which reads $\Gamma_{23}=\Gamma_{12}+\Gamma_{31}.$  With the values of collisional dephasing rates limited by
this relation we observed that our total population always remained
constant. Our results complement with the work in \cite{Be05,Sc04}.

In summary, we have analyzed the interaction of an optical frequency comb with a three-levle $\Lambda$ system that leads to cooling of internal degrees of freedom in the KRb molecule from the Feshbach state into ultracold deeply bound state. Using the reduced density matrix to introduce decoherence in the forms of spontaneous emission and collisional dephasing, we studied the effectiveness of the population transfer into the ultracold state performed by the standard OFC and a sine or cosine-chirped OFC and uncovered a remarkable difference in the system's response. We have shown that in the presence of fast decoherence an efficient cooling of internal degrees of freedom form the Feshbach state may be accomplished by a sinusoidally modulated OFC. While a standard OFC implementation leads to a substantial loss of the efficiency  in the presence of fast dephasing. The sine modulated OFC induces the response in the three-level system that resembles the dark state formed in STIRAP. It negligibly populates the excited state manifold during a stepwise population transfer from the initial to the final, ultracold state and, thus, minimizes the effect of decay and optimally resolves the collisional impact.

 This research is supported by the National Science
Foundation under Grant No. PHY-1205454.

\end{document}